\PassOptionsToPackage{dvipsnames}{xcolor}
\documentclass[longbibliography,nofootinbib,aps,prx,superscriptaddress,twocolumn,10pt]{revtex4-2}

\usepackage{amsmath,amsfonts,amssymb,amsthm,dcolumn,bm,qcircuit,appendix,mathtools,thmtools,thm-restate,mathrsfs,pgfplots,soul,lipsum,graphicx,braket,enumitem,caption,subcaption,ragged2e}
\usepackage[linesnumbered,ruled,vlined]{algorithm2e}

\usepackage{xcolor}
\usepackage{hyperref}
\definecolor{DarkBlueHex}{HTML}{191970} % same as MidnightBlue

\usepackage[capitalize,nameinlink]{cleveref}
\hypersetup{
	colorlinks = true,
	linkcolor = [rgb]{0.70,0.13,0.13},
	citecolor = [rgb]{0.13,0.55,0.13},
	urlcolor = [rgb]{0.25, 0.41, 0.88}}

\usepackage{mathtools}
\usepackage{array}
\usepackage{multirow}
\usepackage{bbold}
\usepackage{ragged2e}%

\usepackage{tikz}
\usetikzlibrary{positioning, shapes.arrows}

\newcolumntype{L}[1]{>{\raggedright\let\newline\\\arraybackslash\hspace{0pt}}m{#1}}
\newcolumntype{C}[1]{>{\centering\let\newline\\\arraybackslash\hspace{0pt}}m{#1}}
\newcolumntype{R}[1]{>{\raggedleft\let\newline\\\arraybackslash\hspace{0pt}}m{#1}}

\numberwithin{proposition}{section}
\numberwithin{theorem}{section}
\numberwithin{lemma}{section}
\numberwithin{corollary}{section}
\numberwithin{remark}{section}
\numberwithin{definition}{section}
\numberwithin{equation}{section}

\DeclareRobustCommand{\orcidicon}{%
	\begin{tikzpicture}
	\draw[lime, fill=lime] (0,0) 
	circle [radius=0.16] 
	node[white] {{\fontfamily{qag}\selectfont \tiny ID}};
	\draw[white, fill=white] (-0.0625,0.095) 
	circle [radius=0.007];
	\end{tikzpicture}
	\hspace{-2mm}
}
\foreach \x in {A, ..., Z}{%
	\expandafter\xdef\csname orcid\x\endcsname{\noexpand\href{https://orcid.org/\csname orcidauthor\x\endcsname}{\noexpand\orcidicon}}
}
\newcommand{\NP}{\textsf{NP}}
\newcommand{\BQP}{\textsf{BQP}}

\begin{document}
\title{Fundamental Limitations of Post-Quantum Cryptographic Architectures}
\author{Jiho Jung\orcidA{}}
\thanks{These authors contributed equally to this work.}
\affiliation{Department of Law, Korean National Police University, Asan 31539, Korea}
\affiliation{Team QST, Seoul National University, Seoul, 08826, Korea}
\author{Donghwa Ji\orcidB{}}
\thanks{These authors contributed equally to this work.}
\affiliation{College of Liberal Studies, Seoul National University, Seoul 08826, Korea}
\affiliation{Team QST, Seoul National University, Seoul, 08826, Korea}
\author{Mingyu Lee\orcidC{}}
\affiliation{Department of Computer Science and Engineering, Seoul National University, Seoul 08826, Korea}
\affiliation{Team QST, Seoul National University, Seoul, 08826, Korea}

\author{Kabgyun Jeong\orcidD{}}
\email{kgjeong6@snu.ac.kr}
\affiliation{Institute of Computer Technology, College of Engineering, Seoul National University, Seoul, 08826, Korea}
\affiliation{Team QST, Seoul National University, Seoul, 08826, Korea}

\date{\today}

\begin{abstract}
Modern lattice-based cryptography, particularly the learning with errors paradigm, relies on injecting artificial noise to secure data against quantum adversaries. This study systematically examines the theoretical and physical boundaries of this noise-reliant model across four interconnected domains: computational complexity, information-theoretic thermodynamics, quantum error correction, and quantum learning theory. Starting from the algorithmic foundation, our analysis notes that these frameworks rely on provisional complexity-theoretic assumptions that remain vulnerable to future quantum algorithmic advancements. Furthermore, by translating this cryptographic mechanism into physical thermodynamics, we illustrate that intentionally injected discrete Gaussian noise does not equate to the permanent erasure of information. Because the structural integrity of the cryptographic secret remains preserved within the ciphertext, advanced quantum error correction protocols and quantum learning models can efficiently extract the underlying mathematical kernel. Ultimately, we suggest that while lattice-based cryptography provides a robust transitional alternative, definitively classifying these frameworks as unconditionally post-quantum represents a premature classification relying on transient physical bottlenecks rather than impenetrable theoretical boundaries.
\end{abstract}
\maketitle

\tableofcontents

%===============
\section{Introduction}

Introduced in 1994, Peter Shor's algorithm fundamentally undermined the RSA cryptosystem by solving the integer factorization problem in polynomial time using the quantum Fourier transform~\cite{Shor99}, prompting the cryptographic community to shift away from algebraically periodic structures toward mathematical problems that offer no obvious quantum advantage. Consequently, lattice-based assumptions, notably the Learning with Errors (LWE) paradigm~\cite{Regev05}, emerged as the foundation for what is now collectively termed post-quantum cryptography, leading institutions such as NIST to rapidly standardize this framework to secure data against future quantum adversaries~\cite{NIST24}.

However, this work suggests that the widely adopted term `post-quantum' creates a profound conceptual ambiguity because the nomenclature implies these cryptosystems are fundamentally immune to the computational power of quantum mechanics, whereas the perceived security of current candidates relies strictly on the empirical observation that efficient quantum attack algorithms have not yet been discovered~\cite{Peikert16}. Rather than possessing a mathematically proven absolute resistance grounded in physical irreversibility, labeling these defensive technologies as unconditionally post-quantum obscures a critical gap between the provisional algorithmic assumptions of these cryptosystems and the underlying physical reality of quantum computation.

To demonstrate this conceptual inadequacy and establish a more rigorous taxonomy, we systematically deconstruct the theoretical and physical boundaries of lattice-based cryptography across four interconnected domains. First, we examine the unresolved boundaries of quantum complexity theory to determine whether the algebraic structure of LWE possesses inherent mathematical immunity to quantum operations. Second, we translate the cryptographic noise mechanism into the rigorous frameworks of information-theoretic thermodynamics~\cite{Shan48, Fano49}. This translation rigorously evaluates the permanence of this informational obfuscation.

Third, we establish the fundamental structural equivalence between lattice cryptographic noise and continuous-variable physical displacement errors. This structural correspondence demonstrates how advanced quantum error correction protocols physically neutralize these intentional perturbations~\cite{Grimsmo21, Noh22}. Finally, we contextualize these vulnerabilities within the rapid advancements of discrete quantum learning models and current noisy intermediate-scale quantum algorithms~\cite{GKZ19, Poremba25, Zeng25}. Ultimately, this comprehensive analysis determines whether the perceived security of this paradigm stems from an impenetrable theoretical lower bound or merely from transient engineering limitations regarding state preparation.

\section{Learning with Errors Problem}

Rigorously evaluating post-quantum cryptography requires formalizing its foundational premise, the learning with errors problem. This section establishes the classical definition of LWE and subsequently reinterprets this problem through the quantum sample model, translating classical cryptographic assumptions into the rigorous language of quantum information theory.

\subsection{The Classical Structure of the LWE Problem}

Introduced by Regev in 2005~\cite{Regev05}, the LWE problem serves as the mathematical bedrock for modern lattice-based cryptographic protocols by injecting artificial, algorithmically generated noise into a system of linear equations to deliberately obstruct classical inversion techniques.

Formally, given an integer modulus $q \ge 2$ and a dimension parameter $n \ge 1$, let the vector $\mathbf{s} \in \mathbb{Z}_q^n$ represent the secret information. The LWE distribution $A_{\mathbf{s}, \chi}$ outputs a sample $(\mathbf{a}, b) \in \mathbb{Z}_q^n \times \mathbb{Z}_q$, where the vector $\mathbf{a}$ is sampled uniformly at random from $\mathbb{Z}_q^n$ and the scalar $b$ is computed as:

$$b = \langle \mathbf{a}, \mathbf{s} \rangle + e \pmod q.$$

Here, $\langle \cdot, \cdot \rangle$ denotes the standard inner product, and the crucial error term $e \in \mathbb{Z}_q$ is drawn from a specific probability distribution $\chi$ over $\mathbb{Z}_q$, typically a discrete Gaussian centered at zero. Unlike random physical thermodynamic noise, this error term $e$ functions as a deterministic mathematical construct explicitly designed to ensure computational obfuscation.

\begin{figure}[h]
    \centering
    \includegraphics[width=\linewidth]{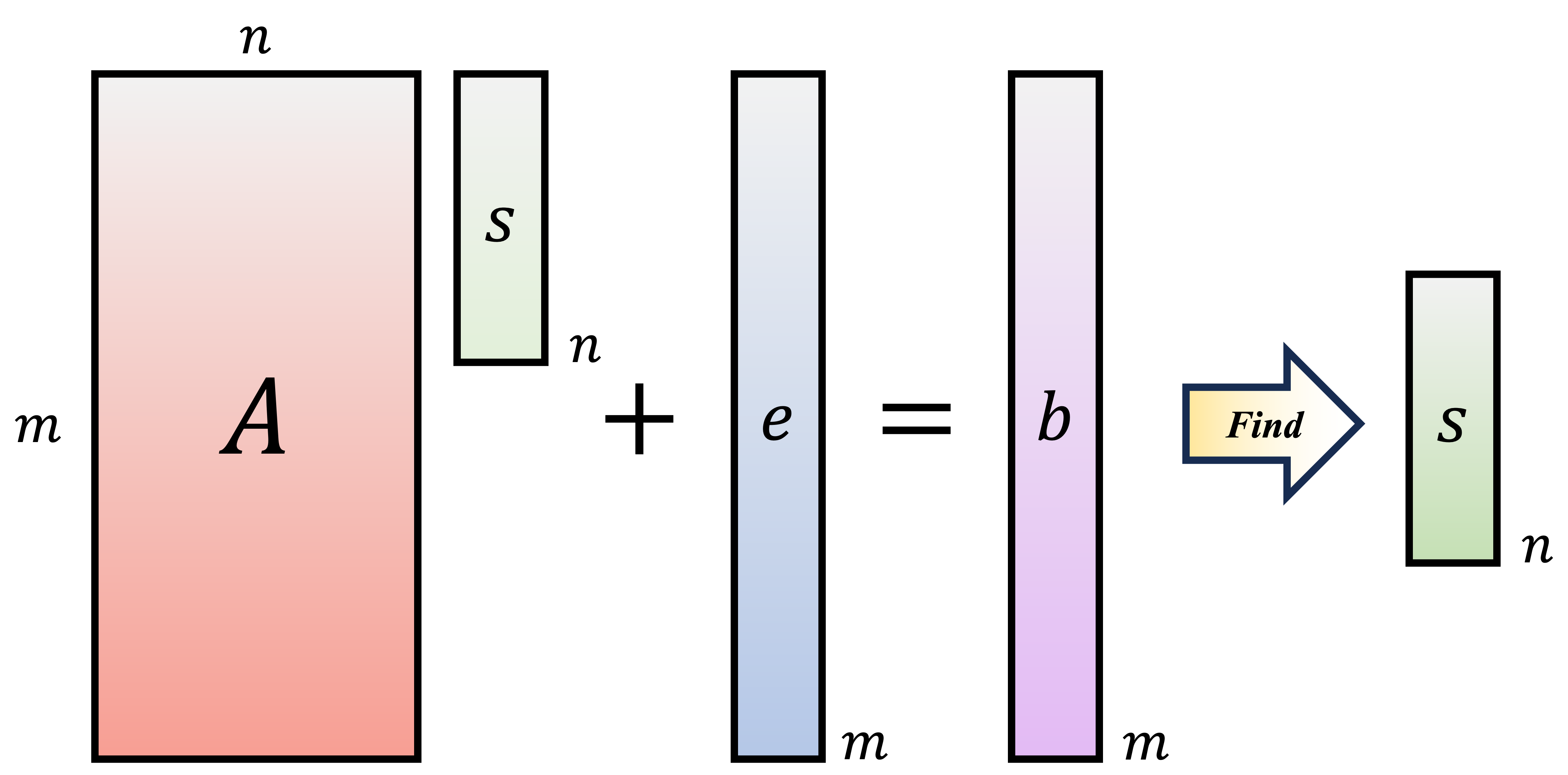}
    \caption{The algebraic structure of the LWE problem, where an artificial error vector $e$ is added to the linear combination of matrix $A$ and secret vector $s$ ($b = A s + e$) to mimic a noisy linear system, rendering the exact inversion of $s$ from $b$ computationally intractable under standard classical algorithms.}
    \label{fig:nlp}
\end{figure}

The LWE problem encompasses two fundamental computational variants:

\begin{itemize}
    \item \textbf{Search LWE:} The task of recovering the exact secret vector $\mathbf{s}$ given access to a polynomial number of samples $(\mathbf{a}_i, b_i)$.
    \item \textbf{Decision LWE:} The task of distinguishing, with a non-negligible advantage, whether a given set of samples is drawn from the LWE distribution $A_{\mathbf{s}, \chi}$ or from a completely uniform distribution over $\mathbb{Z}_q^n \times \mathbb{Z}_q$.
\end{itemize}

While finding $\mathbf{s}$ without the error term $e$ is trivially solvable in polynomial time via Gaussian elimination, introducing the error distribution $\chi$ fundamentally alters the complexity structure. Regev demonstrated that solving LWE is at least as hard as quantumly solving several worst-case lattice problems. Crucially, because LWE is not mathematically proven to reside within the \NP-hard domain, its cryptographic utility relies entirely on presumed average-case intractability supported solely by worst-case reductions.

\subsection{The GKZ Algorithm and Quantum Samples}

Building upon the foundational discovery that quantum learning algorithms demonstrate remarkable robustness against classification noise in parity problems~\cite{Cross15}, evaluating the quantum complexity of LWE requires elevating the classical sample $(\mathbf{a}, b)$ into a quantum mechanical state. The GKZ algorithm utilizes this exact framework~\cite{GKZ19}, demonstrating that LWE becomes efficiently solvable in polynomial time if the adversary possesses access to these quantum samples. Within this model, the quantum sample $|\psi_{\text{LWE}}\rangle$ is not merely a static data structure but a coherent physical superposition over the classical sample space:

$$|\psi_{\text{LWE}}\rangle = \frac{1}{\sqrt{|\mathbb{Z}_q^n|}} \sum_{\mathbf{a} \in \mathbb{Z}_q^n} |\mathbf{a}\rangle |\langle \mathbf{a}, \mathbf{s} \rangle + e_{\mathbf{a}} \pmod q \rangle.$$

Operating conceptually as a noise-tolerant generalization of the Bernstein-Vazirani algorithm~\cite{BV93}, the GKZ algorithm extracts the hidden macroscopic variable $\mathbf{s}$ by applying the quantum Fourier transform to this superposition. This operation translates mathematical obfuscation directly into physical phase dynamics, relying entirely on the coherent manipulation of probability amplitudes. Applying the quantum Fourier transform to the input register $|\mathbf{a}\rangle$ induces a critical phase interference process:

\begin{enumerate}
    \item \textbf{The Ideal Case ($e_{\mathbf{a}}=0$):} Perfect constructive interference occurs, aligning the probability amplitudes perfectly and collapsing the system strictly onto the target basis state $|\mathbf{s}\rangle$ with a probability of $1$ upon projective measurement.
    \item \textbf{The Noisy Case ($e_{\mathbf{a}} \neq 0$):} The cryptographically injected error term $e_{\mathbf{a}}$ functions physically as a random phase shift, threatening to disrupt the interference pattern by inducing artificial decoherence.
\end{enumerate}

Because applying the quantum Fourier transform leaves the target state's probability amplitude tethered to the Fourier coefficients of the error distribution $\chi$, the GKZ formulation rigorously demonstrates that this residual interference signal can be computationally amplified. This amplification allows the original secret information $\mathbf{s}$ to be recovered with overwhelming probability using only a polynomial number of quantum samples. Consequently, the GKZ algorithm exposes that the presumed cryptographic hardness of LWE does not originate from intrinsic mathematical intractability, but rather it is entirely bottlenecked by the technological constraints of the index erasure problem, which governs the physical thermodynamic cost of preparing a coherent computable quantum state from classical data~\cite{QRAM08}.

\subsection{Fault-Tolerant Complexity and the State Preparation Bottleneck}

While the theoretical breakthrough of the GKZ algorithm exposes the mathematical vulnerability of LWE, executing this algorithm requires the adversary to perfectly prepare the massively superposed quantum state $|\psi_{\text{LWE}}\rangle$. This assumption presents a profound physical bottleneck that has historically rendered such quantum attacks practically intractable due to the massive operational costs of Quantum Random Access Memory (QRAM)~\cite{QRAM08}.

Rigorously evaluating this technological barrier requires analyzing the algorithm within the framework of fault-tolerant quantum computation, where the runtime is strictly governed by T-depth complexity, defined as the overall number of non-Clifford gate layers required for execution. Under this paradigm, preparing the full exponential superposition of classical LWE samples demands an astronomical T-depth. 

Furthermore, this preparation phase is strictly bounded by the thermodynamics of the index erasure problem. Loading classical data into a coherent quantum superposition requires reversing classical computational steps, which mandates the erasure of entangled index registers to maintain perfect quantum coherence. According to Landauer's principle~\cite{Landauer61}, this erasure incurs an unavoidable thermodynamic energy cost. Therefore, the absolute security of the LWE framework relies exclusively on this specific operational overhead, and while the mathematical kernel is theoretically solvable, the current cryptographic defense is provided solely by the physical difficulty of loading massive datasets into a fault-tolerant quantum memory.

\section{Theoretical Limitations of the Post-Quantum Paradigm}

This section critically examines the physical and information-theoretic limitations of the so-called `post-quantum' paradigm. Building upon the previously defined LWE and quantum sample models, we deconstruct these limitations from three distinct perspectives.

\subsection{Limitations of the Complexity Assumptions}

This section critically examines the theoretical limitations of the post-quantum cryptographic paradigm by deconstructing its foundational premises, suggesting that the ubiquitous term post-quantum cryptography creates a misleading conceptual framework implying these cryptosystems fundamentally transcend the computational boundaries of quantum mechanics. Quantum key distribution guarantees unconditional information-theoretic security strictly grounded in the fundamental physical laws of quantum mechanics~\cite{Bennett14}. In contrast, the security of prominent post-quantum candidates like lattice-based cryptography arises strictly from the empirical absence of efficient quantum attack algorithms, rather than any mathematically proven intractability or absolute physical law.

Stripped of these absolute physical guarantees, the most fundamental vulnerability of lattice-based cryptography lies within the unresolved boundaries of quantum complexity theory. Classically, the security of the LWE problem is anchored in the assumed average-case hardness of specific worst-case lattice problems. Because there exists no mathematical proof guaranteeing that these specific structured problems reside outside the bounded-error quantum polynomial time (\BQP) class, their cryptographic utility relies entirely on an unproven computational complexity assumption~\cite{Peikert16}.

\begin{figure}[t]
    \centering
    \includegraphics[width=\linewidth]{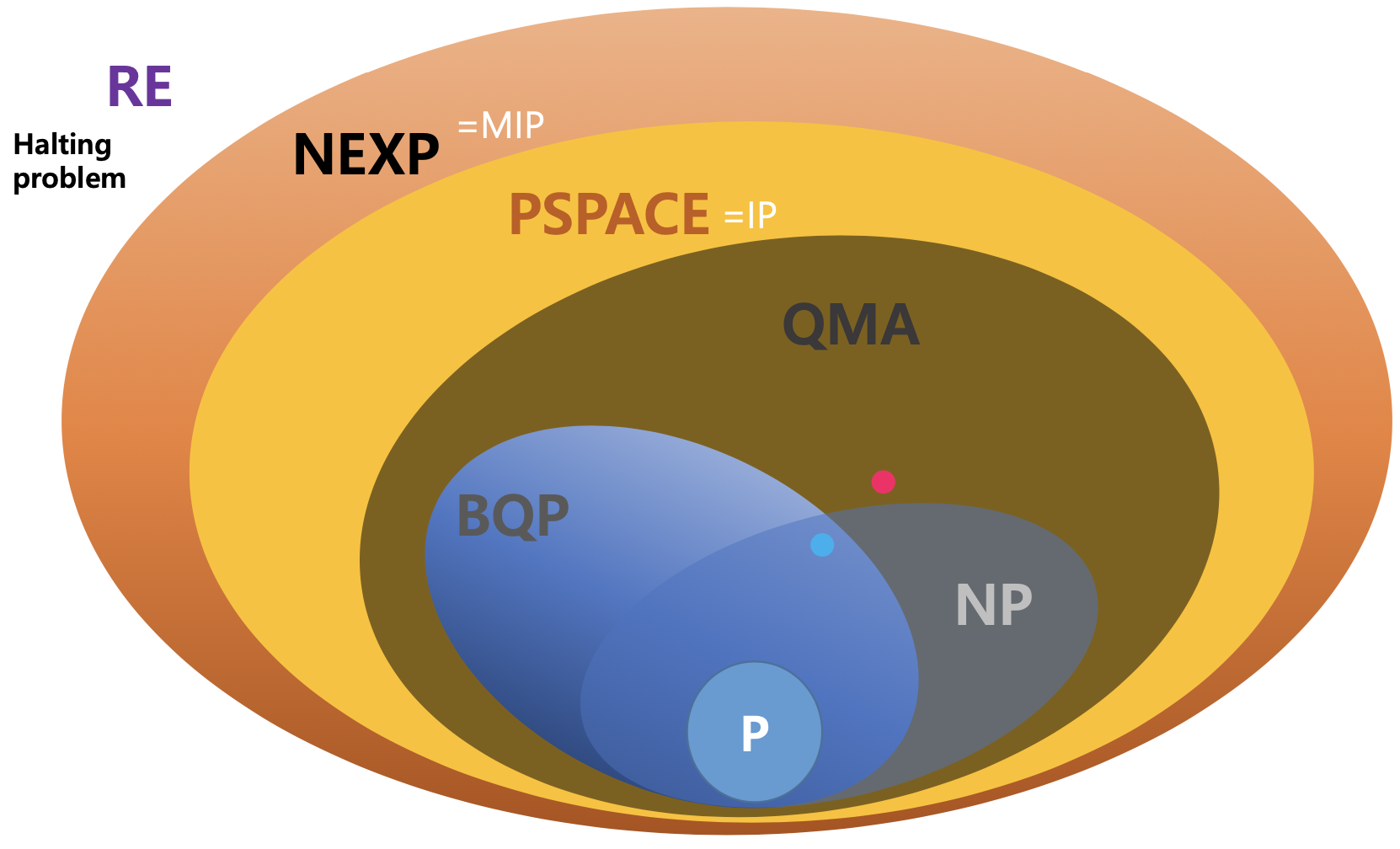}
    \caption{Conceptual diagram of computational complexity classes. The LWE problem belongs to \NP, but it is not \NP-Complete. The entire security paradigm of current post-quantum cryptography relies precariously on the unproven assumption that specific lattice problems do not intersect with the $\BQP$ class.}
    \label{fig:complexity}
\end{figure}

The celebrated reduction by Regev~\cite{Regev05} establishes that solving LWE on average is as difficult as solving standard lattice problems in the worst case. However, this reduction is mathematically valid only when utilizing large polynomial approximation factors. This means that within the specific parameter regimes required for efficient encryption and decryption protocols, the underlying lattice problems are mathematically known to reside far outside the \NP-hard domain. This discrepancy introduces a critical theoretical gap where the operational instances of LWE might fundamentally collapse into the $\BQP$ complexity class without violating any known computational lower bounds.

Compounding this complexity-theoretic fragility, the fundamental algebraic architecture of the LWE problem exhibits precarious structural similarities to mathematical problems already demonstrably compromised by established quantum algorithms. Shor's algorithm efficiently resolves the hidden subgroup problem over finite Abelian groups by utilizing the quantum Fourier transform to extract hidden periodicities~\cite{Shor99}. Consequently, the standard LWE problem can conceptually be interpreted as merely a noisy variant of this exact algebraic structure. 

To defend against this established quantum algebraic framework, the cryptographic community places substantial faith in the assumption that artificially injecting discrete Gaussian noise will successfully diffuse the necessary phase interference patterns. This injected noise is intended to prevent the quantum Fourier transform from isolating the hidden secret. However, as explicitly demonstrated by recent quantum learning models, this noise functions physically merely as a random phase shift rather than an insurmountable computational hurdle~\cite{Cross15, GKZ19}. This establishes that the underlying algebraic structure possesses no inherent mathematical immunity to quantum operations, rendering the 'unconditionally secure' classification premature.

\subsection{Information-Theoretic Conservation of Cryptographic State}

The fundamental security premise of lattice-based cryptography, particularly LWE, relies on obscuring the exact solution by systematically injecting artificial error terms into a linear system. While this error is treated computationally as a definitive algorithmic barrier, translating this mechanism into the rigorous frameworks of quantum information theory reveals a different reality. The introduction of noise does not equate to the irreversible erasure or permanent inaccessibility of information.

\begin{figure*}[t]
    \centering
    \includegraphics[width=\linewidth]{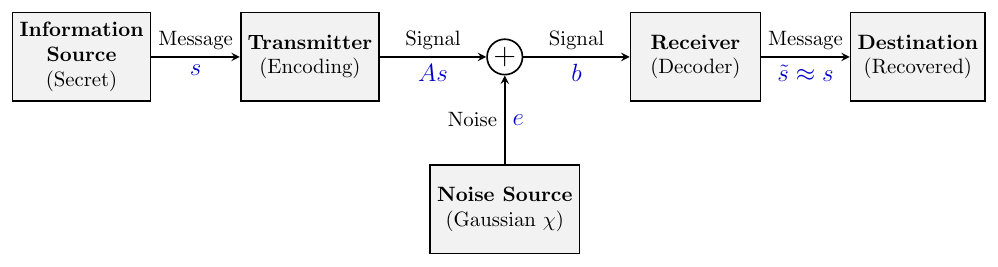}
    \caption{Structural correspondence between Shannon's Communication Model and LWE. While the algorithmic error term ($e$) is designed to obstruct cryptographic inversion, it functions strictly as channel noise from an information-theoretic standpoint. Shannon's theorem dictates that noise within the channel capacity is fundamentally correctable through optimal decoding.}
    \label{fig:shannon_model}
\end{figure*}

The foundation of this critique lies in Shannon's noisy-channel coding theorem~\cite{Shan48}, which fundamentally governs the physical limits of information preservation. Shannon rigorously demonstrated that error-free information transmission remains physically achievable across a noisy channel provided the transmission rate $R$ remains strictly below the channel capacity $C$, defined mathematically by the maximum mutual information 

$$C = \max_{p(x)} I(X; Y)$$

between the transmitted signal $X$ and the received signal $Y$. This fundamental theorem establishes that noise is not an absolute thermodynamic barrier but rather a physical perturbation requiring an engineering energy cost to rectify. Furthermore, because the error distribution $\chi$ injected into the LWE problem typically manifests as a discrete narrow Gaussian distribution centered at zero, this mathematical construct remains statistically indistinguishable from generic thermal noise occurring within physical quantum channels.

Building upon this fundamental channel capacity, the informational preservation of the cryptographic state is naturally extended and strictly quantified through Fano's inequality~\cite{Fano49}. The LWE cryptosystem strictly requires the legitimate receiver to successfully decrypt the message using a secret trapdoor, forcing the overall probability of a decoding error $P_e$ to asymptotically approach zero. This absolute requirement for functional correctness mathematically binds the conditional entropy of the secret $s$ given the public ciphertext $c$ according to the relation:

$$H(s|c) \le H(P_e) + P_e \log(|S| - 1),$$
where $H(P_e)$ represents the binary entropy of the error probability and $|S|$ denotes the size of the secret space. As the error probability vanishes, this strict mathematical bound firmly proves that the original informational content is not thermodynamically annihilated. It is merely distributed across the noise variables, ensuring that the structural integrity of the secret remains fully preserved within the observable ciphertext.

Translating this classical informational bound into the quantum physical regime, a more rigorous proof of this vulnerability is rooted in the fundamental continuity of von Neumann entropy. The difference in informational entropy between the ideal noiseless quantum state $\rho$ encoding the secret and the actual noisy state $\sigma$ corresponding to the LWE sample is tightly bounded by their trace distance $T(\rho, \sigma)$ according to the Fannes--Audenaert inequality~\cite{Aud07}:

$$|S(\rho) - S(\sigma)| \le T \log(d - 1) + H(T),$$
where $d$ is the dimension of the underlying Hilbert space and $H(T)$ is the binary entropy function. The dimension $d$ in continuous quantum spaces can become arbitrarily large and potentially cause the bound to diverge. However, the discrete Gaussian error distribution injected into LWE inherently localizes the quantum state to a finite effective dimension. This structural constraint has been recently operationalized by advanced $\epsilon$-net randomizing techniques that mathematically compress the necessary quantum sample space to bypass thermodynamic bottlenecks~\cite{Jeong23}. 

Because the cryptographically permissible error in LWE must remain sufficiently small to allow classical decryption, the thermodynamic entropy difference $|S(\rho) - S(\sigma)|$ remains fundamentally constrained. Therefore, the post-quantum security of LWE relies entirely on the temporary informational obfuscation of the data rather than its absolute physical absence, constituting a transient state that can be deterministically filtered.

\subsection{Physical Equivalence to Quantum Error Correction}

This informational vulnerability becomes far more acute within the context of quantum error correction. The Gottesman-Kitaev-Preskill (GKP) code plays a foundational role within continuous-variable systems~\cite{GKP01}. This premier bosonic error-correcting architecture is physically engineered to correct environmental displacement errors. It achieves this by embedding discrete quantum information into a continuous periodic lattice structure within phase space. Consequently, the mathematical error term of LWE shares a fundamental structural equivalence with these physical displacement errors.

LWE is strictly defined over a discrete algebraic ring $\mathbb{Z}_q$. However, its injected cryptographic noise essentially functions as a discretized continuous Gaussian distribution. The boundary delineating this discretized mathematical noise from continuous thermodynamic noise is entirely artificial. This functional equivalence allows physicists to implement fault-tolerant quantum error correction protocols. These protocols actively filter Gaussian noise surrounding lattice points. They effectively restore the perturbed state via GKP stabilizing mechanisms. 

The physical mechanics of this restoration perfectly mirror cryptographic decryption. GKP error correction relies on periodic stabilizer measurements~\cite{CampagneIbarcq20}. These measurements extract an error syndrome without collapsing the encoded logical state. The syndrome accurately quantifies the exact continuous shift caused by the noise~\cite{Grimsmo21}. This physical extraction completely isolates the injected LWE error. The underlying structural integrity of the message remains totally undisturbed. Therefore, the cryptographic obfuscation is physically dismantled.

Furthermore, modern fault-tolerant architectures actively address larger error variances. If the injected LWE noise exceeds the fundamental correction bound, it induces a logical error. The state incorrectly shifts to an adjacent lattice point. Advanced quantum architectures solve this vulnerability through code concatenation~\cite{Noh22}. They embed the continuous-variable GKP code within a discrete topological code. The Surface code operates as a primary topological layer in this dual architecture~\cite{Kitaev03}. This hierarchical structure provides exponential error suppression. The GKP layer absorbs the continuous Gaussian drifts. The topological layer concurrently corrects the larger discrete jumps. This guarantees the absolute physical recovery of the underlying cryptographic state.

If a highly coherent quantum computer maps the classical LWE instance into its continuous-variable Hilbert space, the cryptosystem's foundational hard problem changes. It is structurally demoted to a standard and physically correctable displacement error. The rapid evolution of these error correction architectures provides a rigorous counterargument to the assumed security of LWE. The physical techniques developed to identify and correct environmental quantum noise rely entirely on solving bounded distance decoding problems on a lattice. As quantum hardware matures to actively suppress physical errors, the algorithms driving this error correction can be directly redirected. They will systematically neutralize the cryptographic noise of LWE.

\subsection{Algorithmic Extraction via Quantum Learning Models}

Complementing the physical methodologies of quantum error correction, computational learning theory exposes an identical structural vulnerability. Attacking a cryptosystem functions fundamentally as a machine learning task. This task entails the reverse-engineering of a hidden secret function strictly from the statistical correlations between public inputs and encrypted outputs. The LWE problem originally emerged as a hard problem within classical learning theory. This shared theoretical foundation guarantees that LWE possesses structural characteristics perfectly suited for advanced learning algorithms to exploit.

This potential for exploitation is significantly bolstered by the discrete quantum learning paradigm. Certain noisy functions remain strictly unlearnable in polynomial time within the classical probably approximately correct (PAC) learning model~\cite{Valiant84}. However, leveraging coherent quantum superpositions dramatically expands these computational boundaries~\cite{Bshouty95}. If the variance of the error distribution satisfies specific theoretical bounds, the underlying periodicity of the function can be statistically extracted. Under these specific conditions, an efficient quantum learning algorithm inherently exists for the discrete LWE problem~\cite{GKZ19}. 

The sample complexity of this quantum learning algorithm is particularly remarkable. The hidden secret key can be deterministically recovered using merely $O(n \log(1/\eta))$ quantum samples. Furthermore, this inherent quantum learnability equally compromises the conceptually related learning parity with noise problem~\cite{Cross15}. Recent theoretical extensions rigorously prove that advanced quantum algorithms can systematically decrypt even quantum-native noise models. A prime example is the newly defined learning stabilizers with noise problem~\cite{Poremba25}. This continuous structural equivalence proves that lattice-based cryptography remains strictly vulnerable to complete extraction.

Therefore, the assertion that current post-quantum cryptography is unconditionally secure relies entirely on a transient physical bottleneck. However, this final physical bastion is systematically being dismantled. Divide-and-conquer strategies and polynomial T-depth optimizations~\cite{Song22a, Song22b} directly neutralize the computational overhead of state preparation. Advanced $\epsilon$-QRAM architectures linearithmically compress the required sample space~\cite{Jeong23}. 

Crucially, this structural degradation is not strictly confined to the distant realization of fully fault-tolerant architectures. Recent quantum-classical hybrid frameworks and variational quantum algorithms engineered for NISQ devices demonstrate immediate threats. These approaches map the LWE mathematical structure directly into target Hamiltonians~\cite{Zeng25, Zheng25}. By statistically extracting the underlying periodicity through heuristic quantum solvers, these algorithms actively bypass the deep coherent circuit requirements. Ultimately, algorithmic optimizations systematically dissolve the thermodynamic boundaries of quantum state preparation, requiring a rigorous theoretical reassessment of these noise-reliant cryptosystems.

\section{Discussion and Conclusion}

This study has critically examined the ubiquitous classification of lattice-based frameworks as unconditionally post-quantum cryptography. We rigorously evaluated their theoretical and physical boundaries against the operational realities of quantum mechanics. Rather than accepting the conventional premise of mathematical impossibility, our evaluation reveals a different reality. The perceived security of the Learning with Errors paradigm stems primarily from transient engineering limitations regarding coherent noise filtration. It does not originate from an impenetrable theoretical lower bound.

Starting from the algorithmic foundation, our analysis notes that lattice-based cryptography relies heavily on provisional complexity assumptions outside the established \NP-hard domain. Its algebraic architecture exhibits precarious structural similarities to problems already resolved by quantum algorithms. Consequently, its integrity essentially hinges on the practical difficulty of filtering structural noise. Transitioning to quantum information theory, our synthesis of Shannon's theorem and Fano's inequality illustrates a critical point. This synthesis rigorously proves that intentionally injected discrete Gaussian noise does not equate to the permanent thermodynamic erasure of information. Instead, the structural integrity of the cryptographic secret remains fully preserved and mathematically bounded within the ciphertext.

Building directly upon this principle of informational conservation, we demonstrated the physical and algorithmic mechanisms capable of extracting this preserved secret. Physically, the cryptographic error exhibits a fundamental structural equivalence with continuous-variable displacement errors. This equivalence allows advanced quantum error correction protocols to actively detect and restore the perturbed states. Algorithmically, discrete quantum learning models can efficiently dissolve this underlying mathematical kernel. Crucially, this vulnerability is no longer confined to the distant realization of fully fault-tolerant architectures. Hybrid quantum-classical algorithms are already probing these structures within current noisy intermediate-scale quantum environments. Furthermore, recent fault-tolerant optimizations systematically dismantle the thermodynamic overhead previously assumed to protect these cryptosystems.

Lattice-based cryptography undoubtedly provides a robust transitional alternative for the current technological landscape. However, the scientific community must recognize that unconditional security requires the absolute irreversibility of fundamental physical laws. It cannot rely on temporary computational obfuscation. Definitively declaring a post-quantum era based on these evaporating physical bottlenecks represents a premature classification. This reality dictates an immediate need to redefine these noise-dependent alternatives. It also encourages the cryptographic community to continuously test its theoretical boundaries against the relentless advancements of physical implementation.

\section*{Acknowledgements}
We acknowledge helpful discussions with Junseo Lee, Daun Jung, and Yewon Ryu. This work was supported by the National Research Foundation of Korea (NRF) through a grant funded by the Ministry of Science and ICT (Grant Nos. RS-2025-00515537 and RS-2023-00211817). This work was also supported by the Institute for Information \& Communications Technology Promotion (IITP) grant funded by the Korean government (MSIP) (Grant Nos. RS-2019-II190003 and RS-2025-02304540), the National Research Council of Science \& Technology (NST) (Grant No. GTL25011-401), and the Korea Institute of Science and Technology Information (KISTI) (Grant No. P25026). 

\section*{Author Contributions}
K.J. conceived and suggested the main idea. K.J., J.J., and M.L. performed and analyzed the research, and D.H. and M.L. prepared elaborated figures. K.J. supervised the research. All authors wrote and reviewed the manuscript.

\section*{Data Availability}
No datasets were generated or analysed during the current study.

\section*{Conflict of Interest}
The authors declare no conflict of interest.

%%%%%%%%%%%%%%%%%%%%%%%%%%%%%%%%%%%%%%%%%%%%%
\bibliography{reference}

\end{document}